\documentstyle[12pt,epsfig]{article}
\begin{document}

\thispagestyle{empty}

\begin{flushright}
\begin{tabular}{c}
{\bf INFN/BE-00/03} \\ \hline
{\bf 10 Luglio 2000} \\
\end{tabular}
\end{flushright}

\begin{figure}[hp]
\begin{center}
\mbox{\epsfig{file=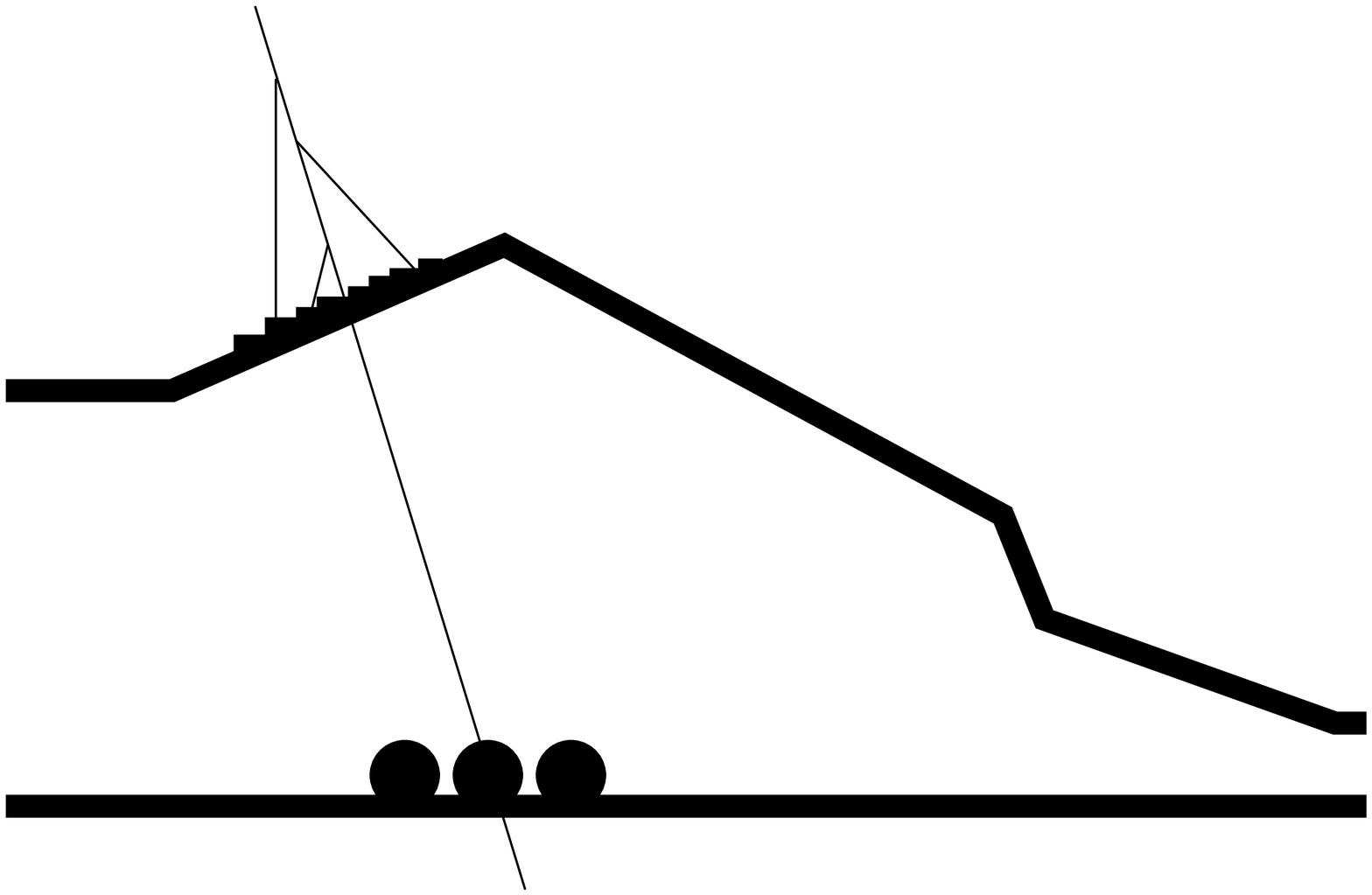,height=10cm,width=15.5cm}}
\end{center}
\end{figure}

\begin{center}
{\Large 
{\bf High sensitivity 2$\beta $ decay study of $^{116}$Cd and $^{100}$Mo}

{\bf with the BOREXINO Counting Test Facility}

{\bf (CAMEO project) }}
\end{center}

\vskip 0.5cm

\begin{center}

G.~Bellini$^{\mbox{a}}$, 
B.~Caccianiga$^{\mbox{a}}$, 
M.~Chen$^{\mbox{c}}$, 
F.A.~Danevich$^{\mbox{b}}$, 
M.G.~Giammarchi$^{\mbox{a}}$,
V.V.~Kobychev$^{\mbox{b}}$, 
B.N.~Kropivyansky$^{\mbox{b}}$, 
E.~Meroni$^{\mbox{a}}$, 
L.~Miramonti$^{\mbox{a}}$, 
A.S.~Nikolayko$^{\mbox{b}}$, 
L.~Oberauer$^{\mbox{d}}$, 
O.A.~Ponkratenko$^{\mbox{b}}$, 
V.I.~Tretyak$^{\mbox{b}}$, 
S.Yu.~Zdesenko$^{\mbox{b}}$, 
Yu.G.~Zdesenko$^{\mbox{b}}$

\vskip 2.0cm

{\Large{\bf INFN - Laboratori Nazionali del Gran Sasso}}\\
\vskip 30pt
\begin{flushright}
\baselineskip 5pt
\footnotesize{\sl {Published by \bf{ SIS-Pubblicazioni}}}\\
\footnotesize{\sl {dei Laboratori Nazionali di Frascati}}
\end{flushright}
\baselineskip 20pt

\end{center}

\newpage
\thispagestyle{empty}
~~~
\newpage
\thispagestyle{empty}

\begin{flushleft}
\begin{tabular}{c}
\footnotesize{{\bf INFN - Istituto Nazionale di Fisica Nucleare}} \\ \hline
\footnotesize{{\bf Laboratori Nazionali del Gran Sasso}} \\ 
\end{tabular}
\end{flushleft}

\begin{flushright}
\begin{tabular}{c}
{\bf INFN/BE-00/03} \\ \hline
{\bf 10 Luglio 2000} \\
\end{tabular}
\end{flushright}

\vskip 4.0cm

\begin{center}

{\Large 
{\bf High sensitivity 2$\beta $ decay study of $^{116}$Cd and $^{100}$Mo}

{\bf with the BOREXINO Counting Test Facility}

{\bf (CAMEO project) }}

\vskip 0.5cm

G.~Bellini$^{\mbox{a}}$, 
B.~Caccianiga$^{\mbox{a}}$, 
M.~Chen$^{\mbox{c}}$, 
F.A.~Danevich$^{\mbox{b}}$, 
M.G.~Giammarchi$^{\mbox{a}}$,
V.V.~Kobychev$^{\mbox{b}}$, 
B.N.~Kropivyansky$^{\mbox{b}}$, 
E.~Meroni$^{\mbox{a}}$, 
L.~Miramonti$^{\mbox{a}}$, 
A.S.~Nikolayko$^{\mbox{b}}$, 
L.~Oberauer$^{\mbox{d}}$, 
O.A.~Ponkratenko$^{\mbox{b}}$, 
V.I.~Tretyak$^{\mbox{b}}$, 
S.Yu.~Zdesenko$^{\mbox{b}}$, 
Yu.G.~Zdesenko$^{\mbox{b}}$

\vskip 0.5cm

\it {
$^{\mbox{a}}$
Physics Dept. of the University and INFN, 20133 Milano, Italy\\
$^{\mbox{b}}$ 
Institute for Nuclear Research, MSP 03680 Kiev, Ukraine\\
$^{\mbox{c}}$
Dept. of Physics, Queen's University, Kingston, Canada\\
$^{\mbox{d}}$
Technical University Munich, Garching, Germany\\
}

\end{center}

\vskip 0.5cm

\begin{abstract}
The unique features (super-low background and large sensitive volume) of the
CTF and BOREXINO set ups are used in the CAMEO project for a high sensitivity
study of $^{100}$Mo and $^{116}$Cd neutrinoless 2$\beta $ decay. Pilot
measurements with $^{116}$Cd and Monte Carlo simulations show that the
sensitivity of the CAMEO experiment (in terms of the half-life limit for $%
0\nu 2\beta $ decay) is (3--5)$\cdot $10$^{24}$ yr with a 1 kg source of $%
^{100}$Mo ($^{116}$Cd, $^{82}$Se, and $^{150}$Nd) and $\approx $10$^{26}$ yr
with 65 kg of enriched $^{116}$CdWO$_4$ crystals placed in the liquid
scintillator of the CTF. The last value corresponds to a limit on
the neutrino mass of $m_\nu \leq 0.06$ eV. Similarly with 1000 kg of $%
^{116}$CdWO$_4$ crystals located in the BOREXINO apparatus the neutrino
mass limit can be pushed down to $m_\nu \leq 0.02$ eV.
\end{abstract}

\newpage
\setcounter{page}{2}

\section{Introduction}

Neutrinoless double beta (0$\nu $2$\beta $) decay is forbidden in the
Standard Model (SM) since it violates lepton number ($L$) conservation.
However many extensions of the SM incorporate $L$ violating interactions and
thus could lead to 0$\nu $2$\beta $ decay \cite{Moe94,Theo98}.
Currently, besides the conventional left-handed neutrino ($\nu $) exchange
mechanism, there are many other possibilities to trigger this process \cite
{Theo98}: right-handed $\nu $ exchange in left-right symmetric models;
exchange of squarks, sneutrinos, etc. via supersymmetric (SUSY)
interactions; exchange of leptoquarks in models with leptoquarks; exchange
of excited Majorana neutrinos in models with composite heavy neutrinos, and
so on. In that sense 0$\nu $2$\beta $ decay has a great conceptual
importance due to the strong statement obtained in a gauge theory of the
weak interaction that a non-vanishing 0$\nu $2$\beta $ decay rate requires
neutrinos to be massive Majorana particles, independently of which mechanism
induces it \cite{Val82}. Therefore, at present 0$\nu $2$\beta $ decay is
considered as a powerful method to test new physical effects beyond the SM,
while absence of this process -- established at the present level of
sensitivity -- would yield strong restrictions on parameters of manifold
extensions of the SM and narrow the wide choice of the theoretical models.
At the same time 0$\nu $2$\beta $ decay is very important in light of the
current status of neutrino physics (see \cite{Neut98}). Indeed, the
solar neutrino problem (in particular lack of $^7$Be neutrinos) and the
deficit of the atmospheric muon neutrino flux \cite{SK98} could be
explained by means of neutrino oscillations, which require nonzero
neutrino masses. Also indication for $\nu _\mu $/$\nu _e$ oscillations was
found by the LSND collaboration \cite{Neut98,LSND95}.
Oscillation experiments are only sensitive to neutrino mass difference,
while measuring the 0$\nu $2$\beta $ decay rate can give the absolute value of
the $\nu $ mass scale, and hence provide a crucial test of neutrino mass
models.

Despite the numerous efforts to observe 0$\nu $2$\beta $ decay beginning
from 1948 up to the present \cite{Moe94} this process still remains
unobserved. The highest half-life limits were set in direct experiments with
several nuclides: $T_{1/2}$(0$\nu )\geq 10^{22}$ yr for $^{82}$Se \cite{Se82}%
, $^{100}$Mo \cite{Mo100}; $T_{1/2}$(0$\nu )\geq 10^{23}$ yr for $^{116}$Cd 
\cite{Dan99,Cd-00}, $^{130}$Te \cite{Te130} and $^{136}$Xe \cite{Xe136}; and
T$_{1/2}$(0$\nu )\geq 10^{25}$ yr for $^{76}$Ge \cite{Ge76,IGEX}.

The present theoretical and experimental status of 2$\beta $ decay 
investigations 
\cite{Moe94,Theo98} makes it necessary to extend the number of candidate
nuclides studied at a sensitivity comparable to 
or better than that for $^{76}$%
Ge (neutrino mass limit $m_\nu $ $\leq 0.1-0.5$ eV). With this aim we
consider in the present paper the use of the super-low background liquid
scintillation detector -- the BOREXINO Counting Test Facility (CTF) \cite
{Bel96} -- for high sensitivity 2$\beta $ decay research.

\section{The CTF and choice of candidate nuclei}

The full description of the CTF and its performance have been published
elsewhere \cite{Bel96,CTF-98,CTF-98A}. Here we recall the main features of
this apparatus, which are important in the following.

The CTF is installed in the Gran Sasso Underground Laboratory and consists
of an external cylindrical water tank ($\oslash $11$\times $10 m; $\approx $%
1000 t of water) serving as passive shielding for 4.8 m$^3$ 
of liquid scintillator
contained in an inner spherical vessel of $\oslash $2.1 m. High purity
water is supplied by the BOREXINO water plant, which provides its
radio-purity level of $\approx 10^{-14}$ g/g (U, Th), $\approx 10^{-10}$ g/g
(K natural) and $<$5 $\mu $Bq/{\it l} for $^{222}$Rn \cite
{Bel96,CTF-98A,Bala96}.

The liquid scintillator is a binary solution of 1.5 g/{\it l} of PPO in
pseudocumene. The fluorescence peak emission is at 365 nm and the yield of
emitted photons is $\approx $10$^4$ per MeV of energy deposited. The
attenuation length is larger than $5$ m above 380 nm \cite{Alimo2000}. The
principal scintillator decay time was measured as $\approx $3.5 ns in a small
volume, while as 4.5--5.0 ns with a source placed in the center of the CTF.
The scintillator is purified by recirculating it from the inner vessel
through a Rn stripping tower, a water extraction unit, 
a Si-Gel column extraction unit,
and a vacuum
distillation unit. It ensures that $^{232}$Th and $^{238}$U contaminations
in the liquid scintillator are less than $(2$--$5)\cdot 10^{-16}$ g/g.

The inner vessel for the liquid scintillator is made of nylon film, 500 $\mu 
$m thick, with excellent optical clarity at 350-500 nm, which allows 
collection of scintillation light with the help of 100 phototubes (PMT) fixed
to a 7 m diameter support structure inside the water tank. The PMTs are 8''
Thorn EMI 9351 tubes made of low radioactivity Schott 8246 glass, and
characterized by high quantum efficiency (26\% at 420 nm), limited transit
time spread ($\sigma $ = $1$ ns), good pulse height resolution for single
photoelectron pulses (Peak/Valley = $2.5$), low dark noise rate (0.5 kHz),
low after pulse probability (2.5\%), and a gain of 10$^7$. The PMTs are
fitted with light concentrators 57 cm long and 50 cm diameter aperture. 
They provide 20\% optical coverage. The number of photoelectrons per MeV measured
experimentally is (300 $\pm $ 30)/MeV on average.
An upgrade of the CTF was realized in 1999 when an additional nylon barrier
against radon convection and a muon veto system were installed.

For each event the charge and timing (precision of 1 ns) of hit PMTs are
recorded. Each channel is doubled by an auxiliary channel to record all
other events coming within a time window of 8 ms after the first event. For
longer delay the computer clock is used (accuracy of $\approx $0.1 s). Event
parameters measured in the CTF include:

- the total charge collected by the PMTs during $0$--$500$ ns (event
energy);

- the tail charge ($48$--$548$ ns) used for pulse shape discrimination;

- PMT timing to reconstruct the event in space (resolution of 10--15
cm);

- the time elapsed between sequential events, used to tag time-correlated
events.

Due to all these measures the CTF is the best super-low background
scintillator of large volume at present. Indeed, the total background rate in
the energy region 250 -- 800 keV (so called ''solar neutrino energy
window'') is about 0.3 counts/yr$\cdot $keV$\cdot $kg and is dominated by
external background from Rn in the shielding water ($\approx $30 mBq/m$^3$
in the region surrounding the inner vessel), while the internal background is
less than 0.01 counts/yr$\cdot $keV$\cdot $kg. Therefore one can conclude
that the CTF is the ideal apparatus for super-low background 2$\beta $ decay
research.

For the choice of 2$\beta $ candidate nuclei let us express the $0\nu 2\beta $
decay probability (neglecting right-handed contributions) as follows \cite
{Theo98}:

\begin{equation}
\left( T_{1/2}^{0\nu }\right) ^{-1}=G_{mm}^{0\nu }\cdot \left| ME\right|
^2\cdot \left\langle m_\nu \right\rangle ^2,
\end{equation}
where $\left| ME\right| $ is the nuclear matrix element of the $0\nu 2\beta $
decay, and $G_{mm}^{0\nu }\left( Z,Q_{\beta \beta }\right) $ is the phase
space factor. Ignoring for the moment the $\left| ME\right| $
calculation \cite{Theo98}, it is evident from Eq. (1) that for the
sensitivity of the $2\beta $ decay study with a particular candidate the most
important parameter is the available energy release ($Q_{\beta \beta }$).
First, because the phase space integral (hence, $0\nu 2\beta $ decay
rate) strongly depends on $Q_{\beta \beta }$ value (roughly as $Q_{\beta
\beta }^5$). Second, the larger the $2\beta $ decay energy, the simpler -- from
an experimental point of view -- to overcome background problems. The crucial
value is 2614 keV which is the energy of the most dangerous $\gamma $'s from 
$^{208}$Tl decay ($^{232}$Th family). Among 35 candidates there are only six
nuclei with $Q_{\beta \beta }$ larger than 2.6 MeV \cite{Aud95}: $^{48}$Ca ($%
Q_{\beta \beta }$ $=4272$ keV, natural abundance $\delta =0.187\%$), $^{82}$%
Se ($Q_{\beta \beta }$ $=2995$ keV, $\delta =8.73\%$), $^{96}$Zr ($Q_{\beta
\beta }$ $=3350$ keV, $\delta =2.80\%$), $^{100}$Mo ($Q_{\beta \beta }$ $%
=3034$ keV, $\delta =9.63\%$), $^{116}$Cd ($Q_{\beta \beta }$ $=2805$ keV, $%
\delta =7.49\%$), and $^{150}$Nd ($Q_{\beta \beta }$ $=3367$ keV, $\delta
=5.64\%$). The values of the phase space integral $G_{mm}^{0\nu }$ for these
candidates are (in units 10$^{-14}$ yr): 6.4 ($^{48}$Ca), 2.8 ($^{82}$Se),
5.7 ($^{96}$Zr), 4.6 ($^{100}$Mo), 4.9 ($^{116}$Cd), and $\approx $20 ($%
^{150}$Nd) \cite{Theo98}. In comparison, $G_{mm}^{0\nu }$ for $^{76}$Ge is
equal $\approx $0.6 (in the same units) because of the lower $2\beta $ decay
energy ($Q_{\beta \beta }$ $=2039$ keV). From this list $^{100}$Mo and $%
^{116}$Cd were chosen as candidates for $2\beta $ decay study with the CTF
in the first phase. The main reason for the 
choice of $^{100}$Mo  -- in addition to its high 
$Q_{\beta \beta }$ value -- is the fact that the Institute for Nuclear Research
(Kiev) possesses $\approx $1 kg of $^{100}$Mo enriched to $\approx $%
99\%. The $^{116}$Cd was chosen because during the last decade the INR (Kiev)
has developed and performed $2\beta $ decay experiments with this nuclide 
\cite{Dan95,Dan98,Dan99,Cd-00}, which can be considered as a pilot
step for the proposed project with the CTF.

\section{CAMEO-I experiment with $^{100}$Mo in the CTF}

There are two different classes of $2\beta $ decay experiments: (a) an
''active'' source experiment, 
in which the detector (containing $2\beta $ candidate nuclei)
serves as source and detector simultaneously; 
(b) an experiment with ''passive'' source
which is introduced in the detector system \cite{Moe94}. The sensitivity of
any $2\beta $ decay apparatus is determined first, by the available
source strengths (mass of the source), and second, by the detector
background. Another factor very essential for determining the
sensitivity is the
energy resolution of the detector. Indeed, for a detector with poor energy
resolution the events from the high energy tail of the $2\nu 2\beta $ decay
distribution run into the energy window of the $0\nu 2\beta $ decay peak,
and therefore generate background which cannot be discriminated from the 
$0\nu 2\beta $ decay signal, even in principle. All of the decay
features are similar: the same two particles are emitted simultaneously from
one point of the source, in the same energy region and with identical
angular distribution. However, the better the
energy resolution -- the smaller
the $2\nu $ tail becomes within the $0\nu $ interval, and thus the
irreducible background becomes lower. Hence, we can conclude that the ultimate
sensitivity to detect $0\nu 2\beta $ decay is really limited by the energy
resolution of the detector, which is the most crucial parameter for any kind
of set up for $2\beta $ decay study.

For the second class of experiments the sensitivity is also restricted
by the trade-off between source strengths and detection efficiency.
The number of $2\beta $ decay candidate nuclei can be enlarged
by increasing the source thickness, which at the same time will lead to 
lower detection efficiency caused by absorption of electrons in the source
and transformation of the measured $2\beta $ decay spectra (broadening of
the peak and shifting it to lower energies).

These experimental considerations are illustrated in Fig. 1, 
where results of model
experiments to study $2\beta $ decay of $^{100}$Mo are presented. The
following assumptions were accepted for simulation: mass of $^{100}$Mo
source is 1 kg ($\approx $6$\cdot 10^{24}$ nuclei of $^{100}$Mo); measuring
time is 5 years; half-life of $^{100}$Mo two neutrino $2\beta $ decay $%
T_{1/2}$(2$\nu 2\beta )$ $=$ $10^{19}$ yr (e.g. see ref. \cite{Mo100-95}),
while for 0$\nu $ mode $T_{1/2}$(0$\nu 2\beta )$ = $10^{24}$ yr. The
simulations were performed with the GEANT3.21 package \cite{GEANT}
and event generator DECAY4 \cite{Decay4}, which describes the initial
kinematics of the events in $\alpha $, $\beta $, and 2$\beta $ decay (how
many particles are emitted, their types, energies, directions and times of
emission). The initial $2\beta $ decay spectra are shown in Fig. 1a and Fig.
1b for different vertical scales. These spectra simulate $%
2\beta $ decay of $^{100}$Mo nuclei placed in an ideal detector (''active''
source technique) with 100\% efficiency for $2\beta $ decay events and with
zero background (energy resolution and energy threshold of 10 keV are
presumed). For the next step the $^{100}$Mo source was introduced in the same
detector as a foil (''passive'' source technique). The
simulated spectra are depicted in Fig. 1c (thickness of $^{100}$Mo foil is
15 mg/cm$^2$) and Fig. 1d (60 mg/cm$^2$). Finally the energy resolution 
of the detector ($FWHM$) was taken into
account and results are shown in Fig. 1e ($FWHM$ = 4\% at 3 MeV) and Fig. 1f
($FWHM$ = 8.8\% at 3 MeV). It should be stressed that Fig. 1 represents the
results of an ideal experiment, while in any real study
the available results can only be worse by reason of the actual background,
higher energy threshold and lower detection efficiency, etc. In fact, this
is very strong statement because it allows to set the sensitivity limit for
any real apparatus. For instance, it is evident from Fig. 1f that 0$\nu
2\beta $ decay of $^{100}$Mo with half-life $T_{1/2}$ = $10^{24}$ yr will 
hardly be observed by using the 
''passive'' source technique with the following
characteristics: (i) product of detection efficiency by number of $^{100}$Mo
nuclei $\approx $6$\cdot 10^{24}$ (e.g., 1 kg of $^{100}$Mo at 100\%
efficiency, or 10 kg of $^{100}$Mo at 10\% efficiency); (ii) $^{100}$Mo
source thickness of 60 mg/cm$^2$; (iii) $FWHM$ = 8.8\% at 3 MeV. At the same
time, it is obvious from Fig. 1e that such a goal can be reached with
similar apparatus with $FWHM$ = 4\% at 3 MeV, and with a 15 mg/cm$%
^2$ source.

\nopagebreak
\begin{figure}[ht]
\begin{center}
\mbox{\epsfig{figure=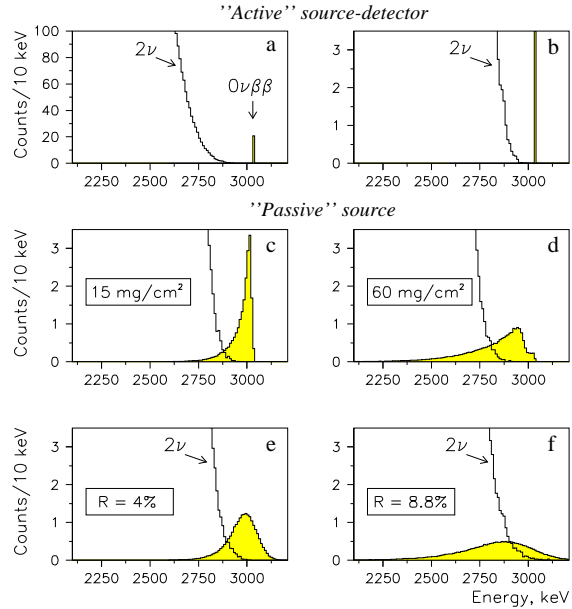,height=8.0cm}}
\caption {Simulated spectra of the model $2\beta $ decay experiment with
1 kg of $^{100}$Mo. (1a, 1b) ''Active'' source technique: $^{100}$Mo nuclei
in a detector with 100\% efficiency, zero background, and with 10 keV
energy resolution. (1c, 1d) ''Passive'' source technique: $^{100}$Mo source
in the same detector with foil thickness 15 mg/cm$^2$
(1c) and 60 mg/cm$^2$ (1d). (1e) The same as (1c) but the energy
resolution ($FWHM$) of the detector at 3 MeV is 4\%.
(1f) The same as (1d) but with $FWHM=8.8$\%.}
\end{center}
\end{figure}

The last solution requires an increase of source area, which is usually limited
by the available dimensions of the low background detectors used for $2\beta 
$ decay search. However, the unique features of the CTF -- large sensitive
volume and super-low background rate of the detector -- permit an 
advanced $2\beta $ decay study of $%
^{100}$Mo with the help of large square ($\approx $7 m$^2$) and thin $^{100}$%
Mo foil placed inside the liquid scintillator.

The $^{100}$Mo source in the CTF is a complex system placed in the inner
vessel with liquid scintillator. It can be represented by three mutually
perpendicular and crossing flat disks with diameter of 180 cm whose centers
are aligned with the center of inner vessel of the CTF. Each disk is
composed of three layers: inner $^{100}$Mo source\footnote{%
In fact, in the plane of the disk the $^{100}$Mo source itself consists of four
sectors with spacing between them of 12 cm, which helps in spatial
reconstruction when events occur near the crossing of the disks.} (thickness
of $\approx $15 mg/cm$^2$) placed between two plastic scintillators of 1 mm
thickness. The inner side of each plastic is coated with thin Al foil serving as
light reflector, while the whole source ''sandwich'' must be encapsulated by
thin transparent film (made of teflon or syndiotactic polypropylene) to
avoid plastic dissolution in liquid scintillator. Plastic detectors can have a
much longer decay constant (f.e. as Bicron plastic BC-444 with 
$\tau \approx 260$ ns 
and light output $\approx $40\% of anthracene) with respect to the
liquid scintillator, thus their pulses can be discriminated easily from the
liquid scintillator signals. The plastics tag electrons
emitted from the $^{100}$Mo source, reducing the background of this
complex detector system significantly. The energy loss measured by the plastics
are added to the electron energy deposit in the liquid scintillator 
to obtain an accurate value of the electron energy and to improve the energy
resolution of the whole detector.

\subsection{Light collection, energy and spatial resolution}

The energy resolution of the scintillation detector depends mainly on the
quality of scintillator itself, the fraction of light collected by PMTs,
uniformity of light collection, quantum efficiency and noise of the PMT
photocathodes, stability and noise of the electronics. The excellent liquid
scintillator used in the CTF yields about 10$^4$ emitted photons per MeV
of energy deposited. In the present CTF design the actual optical coverage
is 20\% and the number of photoelectrons (p.e.) per MeV measured
experimentally is (300 $\pm $ 30)/MeV on average. Thus with fourfold
increase of light collection it would yield $\approx $1200 p.e. per 1 MeV
or $\approx $3600 p.e. for 3 MeV.

To increase the actual optical coverage in the CTF, the PMTs can be mounted
closer to the center of the detector.
For instance, if 200 PMTs are fixed at diameter 5 m (and correspondingly
the light concentrators' entrances at diameter 4 m), or
96 PMTs are fixed at diameter 3.8 m, the optical coverage is equal 
$\approx $80\%. We consider below the last configuration because it is the 
worst case for background contribution from the PMTs\footnote{Special
R\&D is in progress now to find optimal solution for the required 80\%
optical coverage in the CTF.}.
Since the whole volume of the
scintillator is divided by $^{100}$Mo sources into 8 sectors, the PMTs are
split into 8 groups of 12 PMTs each, so that one sector is viewed by one PMT
group. Within the single sector (three mutually perpendicular reflector
plates) scintillation photons would undergo less than 1.5 reflections on 
average before reaching the light concentrator aperture. The Monte Carlo
simulation of the light propagation in such a geometry were performed with
the help of GEANT3.21 program \cite{GEANT}. The emission spectrum and
angular distribution of scintillation photons were added to GEANT code. The
simulation finds that 3 MeV energy deposit would yield $\approx $3700
photoelectrons allowing a measurement of the 
neutrinoless $2\beta $ decay peak of $%
^{100}$Mo ($Q_{\beta \beta }$ $=3034$ keV) with an energy resolution $FWHM$
= 4\%.

This goal can be reached if the non-uniformity of light collection is
corrected by using accurate spatial information about each event;
hence, the spatial reconstruction ability of the CTF has to be enhanced also.
The results of the Monte Carlo simulation prove such a
possibility and show that spatial resolution of $\approx $5--6 cm can be 
obtained with the upgraded CTF\footnote{The value obtained for the spatial 
resolution should be considered as indicative because in this preliminary
phase of study a simplified model for light propagation in the CTF liquid
scintillator has been used.}. 
Primarily this is due to better light collection
(increased by a factor of four). Secondly, it is owing to the spatial
reconstruction method based on the comparison of pulse amplitudes from the
different PMTs (within one group of 12 PMTs), and at the same time due to
analysis of the time structure of each pulse (which can include direct and
reflected light). 

\subsection{Background simulation}

The model of the CAMEO-I set up with $%
^{100}$Mo (described above) is used for the calculations. 
We distinguish here between so called ''$\beta $'' layers
of the liquid scintillator, 15 cm thick\footnote{%
These ''$\beta $'' layers are separated from the total volume of the liquid
scintillator by using the spatial information from the CTF. The thickness of
15 cm is chosen to guarantee the proper spatial reconstruction accuracy and
the full absorption of the electrons emitted in the 2$\beta $ decay of $%
^{100}$Mo.} on both sides of the complex $^{100}$Mo source, and
the rest of the liquid scintillator volume serving as an active shield for
these main inner layers. In such a detector system the following energies
are measured:

i) $E_1^{pl}$ and $E_2^{pl}$ are the energy losses in the first and second
plastic;

ii) $E_1^\beta $ and $E_2^\beta $ are the energy deposits in the first and
second ''$\beta $'' layer;

iii) $E^{ls}$ is the energy loss in the liquid scintillator active shield.

The energy threshold values of the detectors are set as $E_{thr}^{pl}$= 15
keV for the plastic and $E_{thr}^{ls}$ = $E_{thr}^\beta $ = 10 keV for the
liquid scintillator. The energy resolution is 
$FWHM^{pl}=\sqrt{10.8\cdot E^{pl}}$ 
for the plastic and 
$FWHM^\beta=\sqrt{4.8\cdot E^\beta}$ 
for the liquid scintillator ($FWHM$, $E^{pl}$ and $E^\beta$ are in keV). The
latter corresponds to the value of 4\% at 3 MeV. The following cuts are used
in the simulation in order to recognize the double $\beta $ decay events:

i) $E_1^{pl}$ or $E_2^{pl}\geq $ $E_{thr}^{pl}$;

ii) $E_1^{pl}+E_2^{pl}$ $\geq $ 300 keV;

iii) if $E_i^\beta \geq $ $E_{thr}^\beta $, the corresponding $E_i^{pl}$
must be $\geq $ $E_{thr}^{pl}$, necessarily;

iv) $E^{ls}\leq $ $E_{thr}^{ls}$, i. e. there is no signal in the liquid
scintillator active shield.

The simulation of the background and decays of radioactive nuclides in the
installation were performed with the help of GEANT3.21 \cite{GEANT} and the
event generator DECAY4 \cite{Decay4}.

\subsubsection{Two neutrino 2$\beta $ decays of $^{100}$Mo}

The half-life of 2$\nu $2$\beta $ decay of $^{100}$Mo has been already
measured as $\approx $10$^{19}$ yr (e.g. ref. \cite{Moe94,Mo100,Mo100-95}),
hence the corresponding activity of a 1 kg $^{100}$Mo source equals $%
\approx $13.2 mBq. The response functions of the CAMEO-I set up for $2\nu
2\beta $ decay of $^{100}$Mo with $T_{1/2}$ $=$ 10$^{19}$ yr as well as for $%
0\nu 2\beta $ decay with $T_{1/2}$ $=$ 10$^{24}$ yr (for comparison) were
simulated as described above, and results are depicted in Fig. 2a. The
calculated values of efficiency for the neutrinoless channel are 80\%
(within energy window 2.8 -- 3.15 MeV), 74\% (2.85 -- 3.15 MeV), and 63.5\%
(2.9 -- 3.15 MeV). Background due to $2\nu 2\beta $ decay distribution are
19.7 counts (2.8 -- 3.15 MeV), 6.1 counts (2.85 -- 3.15 MeV), and 1.3 counts
(2.9 -- 3.15 MeV) for 5 years measuring period.

\nopagebreak
\begin{figure}[ht]
\begin{center}
\mbox{\epsfig{figure=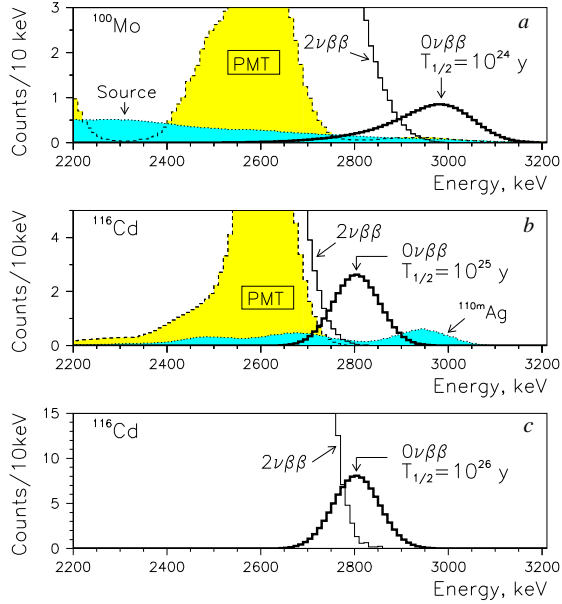,height=8.0cm}}
\caption {(a) The response functions of the CTF (5 kg$\cdot$yr
statistics) for 2$\beta$ decay of $^{100}$Mo with
$T_{1/2}(2\nu)=10^{19}$ yr and
$T_{1/2}(0\nu)=10^{24}$ yr (solid histogram). Total simulated contributions
due to $^{100}$Mo contamination by $^{214}$Bi and $^{208}$Tl (dotted
line), and from $^{214}$Bi and $^{208}$Tl in the PMTs (dashed histogram).
(b) The response functions of the CTF with 65 kg of $^{116}$CdWO$_4$
crystals (5 yr measuring period) for
$2\nu 2\beta $ decay of $^{116}$Cd ($T_{1/2}$ $=$ 2.7$\cdot $10$^{19}$
yr), and $0\nu 2\beta $ decay with $T_{1/2}$ $=$ 10$^{25}$ yr (solid
histogram). The simulated contribution from $^{208}$Tl
in the PMTs (dotted line) and from cosmogenic
$^{110m}$Ag (dashed histogram).
(c) The response functions of the BOREXINO
set up with 1 t of $^{116}$CdWO$_4$ crystals (10 yr measuring time)
for $2\nu 2\beta $ decay of $^{116}$Cd
with $T_{1/2}$ $=$ 2.7$\cdot $10$^{19}$ yr, and $0\nu 2\beta $ decay with $%
T_{1/2}$ $=$ 10$^{26}$ yr (solid histogram).}
\end{center}
\end{figure}

\subsubsection{Radioactive contamination of the $^{100}$Mo source}

The $\approx $1 kg sample of metallic molybdenum enriched in $^{100}$Mo to $%
\approx $99\% -- which has to be applied in the present project -- was
already used in the quest for $2\beta $ decay of $^{100}$Mo to the excited
states of $^{100}$Ru \cite{Mo-92}. In that experiment the radioactive
impurities of the $^{100}$Mo source by $^{40}$K and nuclides from $^{232}$Th
and $^{238}$U chains were measured. Only $^{208}$Tl (measured activity is $%
\leq $ 0.5 mBq/kg) and $^{214}$Bi (measured activity is (12$\pm $3) mBq/kg)
-- due to their high energy release -- could generate the background in the $%
0\nu 2\beta $ decay energy window for $^{100}$Mo. The background problem
associated with $^{100}$Mo radioactive contamination was carefully
investigated by the NEMO collaboration, which is going to begin $2\beta $
decay measurements with $\approx $10 kg of $^{100}$Mo \cite{NEMO98}. It was
found that for the NEMO-3 detector the
maximum acceptable internal activities of $%
^{100}$Mo are 0.3 mBq/kg for $^{214}$Bi and 0.02 mBq/kg for $^{208}$Tl \cite
{NEMO98}. The intensive R\&D were performed by the NEMO collaboration 
with an aim to show that these severe requirements can be reached by using 
presently available physical and chemical methods for 
$^{100}$Mo purification \cite
{Mo-pur98}. On this basis, in our calculation the $^{100}$Mo contamination
criterion for $^{214}$Bi has been taken as 0.3 mBq/kg, and for $^{208}$%
Tl as 0.1 mBq/kg, which is only 5 times better than the actual activity limit in
our $^{100}$Mo sample (0.5 mBq/kg)\footnote{%
We have accepted the less severe and more realistic criterion for $^{208}$%
Tl, because it was shown that chemical purification of Mo is very successful
concerning $^{226}$Ra chain impurities, while for $^{208}$Tl the procedure
is very difficult and less effective \cite{Mo-pur98}.}.

The results of simulation, performed as described above, are as following:
i) $^{214}$Bi contribution to background within the energy interval 2.9 -- 3.15
MeV is 6.5 counts/yr$\cdot $kg; ii) $^{208}$Tl contribution is equal to 0.06
counts/yr$\cdot $kg. The mentioned impurities can be really dangerous
for the experiment. However, there exists a possibility to reduce these
background substantially by using information on the arrival time of each
event for analysis and selection of some decay chains in Th and U families 
\cite{Dan95,Dan96}. With this aim, let us consider the $^{226}$Ra chain
containing $^{214}$Bi: $^{222}$Rn ($T_{1/2}$ = 3.82 d; $Q_\alpha $ = $5.59$
MeV) $\rightarrow $ $^{218}$Po (3.10 m; $Q_\alpha $ = $6.11$ MeV) $%
\rightarrow $ $^{214}$Pb (26.8 m; $Q_\beta $ = $1.02$ MeV) $\rightarrow $ $%
^{214}$Bi (19.9 m; $Q_\beta $ = $3.27$ MeV) $\rightarrow $ $^{214}$Po (164.3 
$\mu $s; $Q_\alpha $ = $7.83$ MeV) $\rightarrow $ $^{210}$Pb. The great
advantage of the CAMEO-I experiment is the very thin $^{100}$Mo source ($%
\approx $15 mg/cm$^2$), which allows detection of most of $\alpha $ and $%
\beta $ particles emitted before or after $^{214}$Bi decay, and tags the
latter with the help of time analysis of the measured events. Indeed, our
calculation gives the following values of the detection efficiencies: $%
\varepsilon _1$ = 55\% for $^{214}$Po ($\alpha $ particles); $\varepsilon _2$
= 80\% for $^{214}$Pb ($\beta $); $\varepsilon _3$ = 37\% for $^{218}$Po ($%
\alpha $); $\varepsilon _4$ = 32\% for $^{222}$Rn ($\alpha $). The
probability to detect at least one of these decays ($^{214}$Po or $^{214}$Pb
or $^{218}$Po or $^{222}$Rn) can be expressed as:

\begin{center}
$\varepsilon =1-(1-\varepsilon _1)\cdot (1-\varepsilon _2)\cdot
(1-\varepsilon _3)\cdot (1-\varepsilon _4).$
\end{center}

By substituting in this formula the calculated efficiency values, it yields $%
\varepsilon $ = 96.1\%, which means that only $\approx $4\% of the $^{214}$%
Bi decays would not be tagged, i.e. $^{214}$Bi contribution to background
can be reduced by a factor of 25 (to the value of $\approx $0.26 counts/yr$%
\cdot $kg). The expected total $\alpha $ decay rate from $^{238}$U and $^{232}$%
Th families in the entire $^{100}$Mo source is $\approx $300 decays/day,
however for an area of 10$\times 10$ cm it is only 0.4 decays/day, which
allows use of the chains with half-life of 26.8 and 19.9 minutes for time
analysis. 
The simulated background spectrum from the internal $^{100}$Mo source
contamination by the $^{214}$Bi and $^{208}$Tl is presented in Fig. 2a,
where the total internal background rate in the energy interval 2.9 -- 3.15
MeV is 0.3 counts/yr$\cdot $kg or $\approx $1.5 counts for 5 years measuring
period.

\subsubsection{Cosmogenic activities in $^{100}$Mo source}

To estimate the cosmogenic activity produced in the $^{100}$Mo foil, we used
the program COSMO \cite{Mar92} which calculates the production of all
radionuclides with half-lives in the range from 25 days to 5 million years
by nucleon-induced reactions in a given target. This program takes into
account the variation of spallation, evaporation, fission and peripheral
reaction cross sections with nucleon energy, target and product charge and
mass numbers, as well as the energy spectrum of cosmic ray nucleons near the
Earth's surface.

For the CAMEO-I project cosmogenic activities were calculated for $^{100}$Mo
source enriched in $^{100}$Mo to 98.5\% (other Mo isotopes: $^{98}$Mo --
0.7\%, $^{97}$Mo -- 0.1\%, $^{96}$Mo -- 0.2\%, $^{95}$Mo -- 0.2\%, $^{94}$Mo
-- 0.1\%, $^{92}$Mo -- 0.2\%). It was assumed 5 years exposure period and
deactivation time of about one year in the underground laboratory. The
calculation shows that among several nuclides with $T_{1/2}\geq 25$ d
produced in $^{100}$Mo source only two can give some background in the
energy window of the $^{100}$Mo neutrinoless $2\beta $ decay. These are $%
^{88}$Y ($Q_{EC}$=3.62 MeV; $T_{1/2}$=107 d) and $^{60}$Co ($Q_\beta $=2.82
MeV; $T_{1/2}$=5.3 yr). Fortunately their activities are very low ($\approx $%
190 decays/yr for $^{88}$Y and $\approx $50 decays/yr for $^{60}$Co), thus the
estimated background in the energy region of 2.7 -- 3.2 MeV is practically
negligible: $\leq $ 0.02 counts/yr$\cdot $kg from $^{88}$Y activity and $%
\leq $ 0.005 counts/yr$\cdot $kg from $^{60}$Co.

\subsubsection{External background}

There are several origins of the external background in the CAMEO-I experiment,
for example, neutrons and $\gamma $ quanta from natural environmental
radioactivity (mainly from concrete walls of the Gran Sasso Underground
Laboratory), contamination of PMTs by $^{40}$K and nuclides from U and Th
families, Rn impurities in the shielding water, cosmic muons ($\mu $ showers
and muon induced neutrons, inelastic scattering and capture of muons), etc.
From all of them only $\gamma $ quanta caused by PMT contamination and by
Rn impurities in the shielding water were simulated in the present work,
while others were simply estimated as negligible on the basis of the results
of ref. \cite{Dark98}, where such origins and contributions for the GENIUS
project \cite{GENIUS-98} were investigated carefully.

The radioactivity values of the EMI 9351 PMT accepted for the simulation
are: 0.194 Bq/PMT ($^{208}$Tl); 1.383 Bq/PMT ($^{214}$Bi); and 191 Bq/PMT ($%
^{40}$K) \cite{CTF-98A,Xe-CTF}. Also possible $^{222}$Rn activity $\approx $%
30 mBq/m$^3$ in the shielding water (in the region surrounding the inner
vessel) is taken into account. The model of the CAMEO-I detector system
described above was used in the calculations, but with two differences: i) ''%
$\beta $'' layers are considered as liquid
scintillator blocks with dimensions 10$\times $10$\times $10 cm$^3$; ii) the
threshold of the liquid scintillator active shield is increased to 30 keV.
The simulation performed under these assumptions gives the following
background rate in the $0\nu 2\beta $ decay energy interval (2.9 -- 3.15
MeV): i) 0.32 counts/yr$\cdot $kg due to $^{214}$Bi in PMT; ii) practically
zero rates from $^{208}$Tl in the PMTs and $^{222}$Rn in the shielding water.
The total simulated background contributions due to $^{214}$Bi and $^{208}$Tl
contamination of the PMTs is shown in Fig. 2a also. After 5 years it yields 
$\approx $1.6 counts in the energy window 2.9 -- 3.15 MeV. Summarizing all
background sources for 5 years of measurements, one can obtain the total
number of $\approx $4.4 counts in the energy range 2.9 -- 3.15 MeV.

\subsection{Sensitivity of the CAMEO-I experiment with $^{100}$Mo}

The sensitivity of the proposed experiment can be expressed in the term of a
lower half-life limit for the $0\nu 2\beta $ decay of $^{100}$Mo as
following:

\begin{equation}
T_{1/2}\geq \ln 2\cdot N\cdot \eta \cdot t/S,
\end{equation}
where $N$ is the number of $^{100}$Mo nuclei ($\approx $6$\cdot $10$^{24}$
in our case); $t$ is the measuring time (5 years); $\eta $ is the detection
efficiency (63.5\%); and $S$ is the number of effect's events, which can be
excluded with a given confidence level on the basis of measured data. Thus for
the five years CAMEO-I experiment $T_{1/2}$ $\geq $ $($13/$S$)$\cdot $10$%
^{24}$ yr. Taking into account the expected background of 4.4 counts, we can
accept 3--5 events as the value for $S$ (depending on the method of 
estimating $S$  
\cite{PDG96,PDG98}) which gives $T_{1/2}$ $\geq $ (3--5)$\cdot $10$^{24}$
yr and, in accordance with \cite{Klap90}, the limit on the neutrino mass
$<m_\nu >\leq 0.5$ eV. 
On the other hand, it is evident from Fig. 2a that neutrinoless $%
2\beta $ decay of $^{100}$Mo with half-life $T_{1/2}$ = 10$^{24}$ yr can 
certainly be registered: the signal (13 counts) to background (4.4 counts) 
ratio is approximately 3:1.

Similar limits $T_{1/2}(0\nu 2\beta )$ $\geq $ (3--5)$\cdot $10$^{24}$
yr can be obtained by the CAMEO-I set up with other nuclides, $^{82}$Se ($%
Q_{\beta \beta }$ $=2996$ keV), $^{96}$Zr ($Q_{\beta \beta }$ $=3350$ keV), $%
^{116}$Cd ($Q_{\beta \beta }$ $=2804$ keV), and $^{150}$Nd ($Q_{\beta \beta
} $ $=3368$ keV)\footnote{%
We do not include $^{48}$Ca ($Q_{\beta \beta }$ $=4272$ keV) in that list
because of its very low natural abundance ($0.187\%$), and hence extremely
high cost of a one kg $^{48}$Ca source.}. Due to its reasonable cost $^{116}$Cd
is the preferable second candidate after $^{100}$Mo. Note, however, that a
half-life limit of $\approx $5$\cdot $10$^{24}$ yr for $^{150}$Nd would lead
-- on the basis of the nuclear matrix elements calculation \cite{Klap90} --
to a restriction on the neutrino mass $<m_\nu >\leq 0.08$ eV.

\section{High sensitivity $2\beta $ decay study of $^{116}$Cd with the CTF}

The most sensitive $0\nu 2\beta $ results are obtained by using an
''active'' source technique \cite{Moe94}. We recall the highest limits $%
T_{1/2}^{0\nu }$ $\geq $ (1--2)$\cdot 10^{25}$ yr established for $^{76}$Ge
with the help of high purity (HP) enriched $^{76}$Ge detectors \cite
{Ge76,IGEX}, and bounds $T_{1/2}^{0\nu }$ $\geq $ $\sim $1$0^{23}$ yr set
for $^{136}$Xe with a high pressure Xe TPC \cite{Xe136}, $^{130}$Te with TeO$%
_2 $ low temperature bolometers \cite{Te130}, and $^{116}$Cd with $^{116}$%
CdWO$_4$ scintillators \cite{Cd-00}.

Continuing this line, we propose to advance the experiment with $^{116}$CdWO$%
_4$ to the sensitivity level of $\approx $10$^{26}$ yr by exploiting the
advantages of the CTF. The idea is to place $\approx $65 kg of enriched $%
^{116}$CdWO$_4$ crystals in the liquid scintillator of the CTF, which would
be used as a light guide and anticoincidence shield for the main $^{116}$CdWO$%
_4$ detectors (CAMEO-II project). 
To prove the feasibility of this task we are considering in the following 
discussion a pilot $^{116}$Cd experiment, and then the design concept of the
present proposal, as well as problems concerning the light collection,
energy and spatial resolution, background sources and sensitivity estimates
of the CAMEO-II project with $^{116}$Cd.

\subsection{The pilot $^{116}$Cd study}

Here we briefly recall the main results of $^{116}$Cd research performed
during the last decade by the INR (Kiev)\footnote{%
From 1998 this experiment was carried out by the Kiev-Firenze collaboration 
\cite{Cd-00}.} in the Solotvina Underground Laboratory (in a salt mine 430 m
underground \cite{Zde87}), and published elsewhere \cite
{Dan95,Dan98,Dan99,Cd-00}. The cadmium tungstate crystal scintillators,
enriched in $^{116}$Cd to 83\%, were grown for research \cite{Dan95}. The
light output of this scintillator is relatively large: $\approx $40\% of
NaI(Tl) \cite{Geo96}. The refractive index is 2.3.
The fluorescence peak emission is at 480 nm with
principal decay time of $\approx $14 $\mu $s \cite{Faz98}. 
The density of CdWO$_4$ crystal is 7.9 g/cm$^3$, and the 
material is non-hygroscopic and chemically inert.

In the first phase of the study only one $^{116}$CdWO$_4$ crystal (15.2 cm$^3$)
was placed inside a veto plastic scintillator and viewed by a PMT through a
light-guide 51 cm long. Outer passive shielding consisted of 
high-purity copper (5 cm),
lead (23 cm) and polyethylene (16 cm). The background rate in the energy
range 2.7--2.9 MeV ($Q_{2\beta }$=2805 keV) was $\approx $0.6 counts/yr$%
\cdot $kg$\cdot $keV \cite{Dan95}. With 19175 h statistics the half-life
limit for 0$\nu $2$\beta $ decay of $^{116}$Cd was set as $T_{1/2}$(0$\nu $) 
$\geq $ 3.2$\cdot $1$0^{22}$ yr (90\% C.L.) \cite{Dan99}, while for
neutrinoless 2$\beta $ decay with emission of one (M1) or two (M2) Majorons
as $T_{1/2}$(0$\nu $M1) $\geq $ 1.2$\cdot 10^{21}$ yr and $T_{1/2}$(0$\nu $%
M2) $\geq $ 2.6$\cdot 10^{20}$ yr (90\% C.L.) \cite{Dan98}.

In 1998 a new set up with four $^{116}$CdWO$_4$ crystals (total mass 339 g)
was mounted in the Solotvina Laboratory. The enriched detectors are viewed
by a special low background 5'' EMI tube (with RbCs photocathode) through
one light-guide ($\oslash $10$\times $55 cm), which is composed of two glued
parts: quartz 25 cm long and plastic scintillator (Bicron BC-412) 30 cm
long. The main detectors are surrounded by an active shield made of 15
natural CdWO$_4$ crystals of large volume \cite{Geo96} (total mass 20.6 kg).
These are viewed by a low background PMT (FEU-125) through an active
plastic light-guide ($\oslash $17$\times $49 cm). The whole CdWO$_4$ array
is situated within an additional active shield made of plastic scintillator
40$\times $40$\times $95 cm, thus, together with active light-guides,
complete 4$\pi $ active shielding of the main $^{116}$CdWO$_4$ detectors is
provided. The outer passive shield consists of high-purity 
copper ($3$--$6$ cm), lead
($22.5$--$30$ cm) and polyethylene (16 cm). The set up is isolated carefully
against penetration of air which could be contaminated by radon.

The multichannel event-by-event data acquisition is based on two IBM
personal computers (PC) and a CAMAC crate. For each
event the following information is stored on the hard disk of the first PC:
the amplitude (energy), arrival time and additional tags. The second
computer records the pulse shape of the $^{116}$CdWO$_4$ scintillators in
the energy range $0.25$--$5$ MeV. It is based on a fast 12 bit ADC (AD9022)
and is connected to the PC by a parallel digital I/O board (PC-DIO-24 from
National Instruments) \cite{Faz98}. Two PC-DIO-24 boards are used to link
both computers and establish -- with the help of proper software -- a
one-to-one correspondence between the pulse shape data recorded by the
second computer and the information stored in the first PC.

The energy scale and resolution of the main detector -- four enriched
crystals taken as a whole -- were measured with different sources ($^{22}$%
Na, $^{40}$K, $^{60}$Co, $^{137}$Cs, $^{207}$Bi, $^{226}$Ra, $^{232}$Th and $%
^{241}$Am) as $FWHM($keV) = $-44+\sqrt{23.4\cdot E+2773}$, where energy $E$
is in keV. In particular, the energy resolution is equal to 11.5\% at 1064 keV
and 8.0\% at 2615 keV. Also the relative light yield and energy resolution
for $\alpha $ particles were determined as following: $\alpha /\beta
=0.12+1.1\cdot $1$0^{-5}E_\alpha $ and $FWHM_\alpha ($keV) = $0.033E_\alpha $
($E_\alpha $ is in keV). The routine calibration was carried out weekly 
with a $%
^{207}$Bi source and once per two weeks with $^{232}$Th. The dead time was
monitored permanently by an LED optically connected to the main PMT (actual
value $\approx $4.2$\%$).

The background spectrum measured during 4629 h in the new set up with four $%
^{116}$CdWO$_4$ crystals \cite{Cd-00} is given in Fig. 3, where old data
obtained with one $^{116}$CdWO$_4$ crystal of 121 g are also shown for
comparison. The background is lower in the whole energy range, except
for the $\beta $ spectrum of $^{113}$Cd ($Q_\beta $ = $316$ keV) \footnote{%
Abundance of $^{113}$Cd in enriched $^{116}$CdWO$_4$ crystals is $\approx $%
2\% \cite{Dan95}.}. In the energy region $2.5$--$3.2$ MeV (location of
expected 0$\nu $2$\beta $ peak) the background rate is 0.03 counts/yr$\cdot $%
kg$\cdot $keV, twenty times lower than before. It was achieved first, due to
new passive and active shielding, and secondly, as a result of the
time-amplitude and pulse-shape analysis of the data.

\nopagebreak
\begin{figure}[ht]
\begin{center}
\mbox{\epsfig{figure=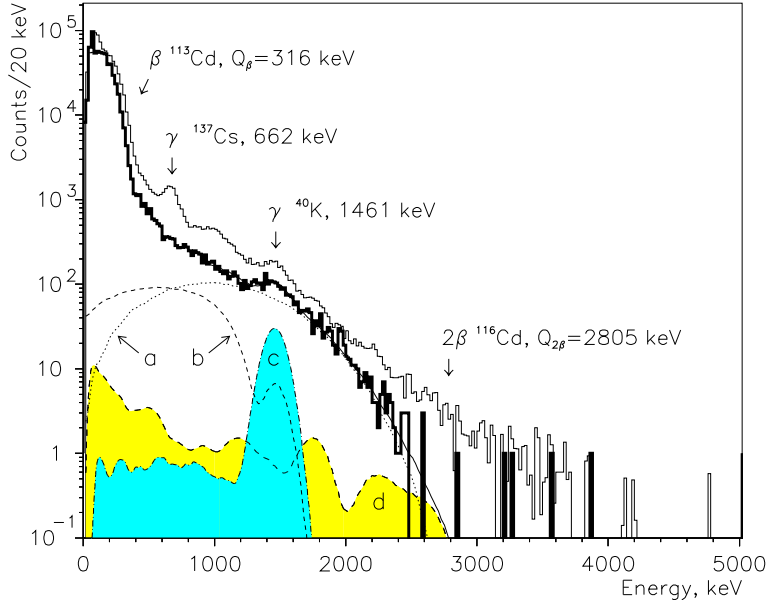,height=8.0cm}}
\caption {Background spectrum of four enriched $^{116}$CdWO$_4$ crystals
(339 g) measured during 4629 h (solid histogram). The old data
with one $^{116}$CdWO$_4$ crystal (121 g; 19986 h) normalized to 339 g and
4629 h (thin histogram). The model components:
(a) 2$\nu $2$\beta $ decay of $^{116}$Cd (fit
value is $T_{1/2}$(2$\nu )=$ 2.6(1)$\cdot $1$0^{19}$ yr); (b) $^{40}$K in
the $^{116}$CdWO$_4$ detectors ($0.8\pm0.2$ mBq/kg);
(c) $^{40}$K in the shielding
CdWO$_4$ crystals ($2.1\pm0.3$ mBq/kg); (d) $^{226}$Ra and $^{232}$Th in the
PMTs.}
\end{center}
\end{figure}

As an example of the time-amplitude technique we consider here in detail the
analysis of the following sequence of $\alpha $ decays from the $^{232}$Th
family: $^{220}$Rn ($Q_\alpha $ = $6.40$ MeV, $T_{1/2}$ = $55.6$ s) $%
\rightarrow $ $^{216}$Po ($Q_\alpha $ = $6.91$ MeV, $T_{1/2}$ = $0.145$ s) $%
\rightarrow $ $^{212}$Pb. The electron equivalent energy for 
$^{220}$Rn $\alpha $ particles in 
$^{116}$CdWO$_4$ is $\approx $1.2 MeV, thus all
events in the energy region $0.7$--$1.8$ MeV were used as triggers. Then all
signals following the triggers in the time interval $10$--$1000$ ms (94.5\%
of $^{216}$Po decays) were selected. The spectra of the $^{220}$Rn and $%
^{216}$Po $\alpha $ decays obtained in this way from data -- as well as the
distribution of the time intervals between the first and second events --
are in an excellent agreement with those expected from $\alpha $ particles
of $^{220}$Rn and $^{216}$Po \cite{Cd-00}. Using these results and taking
into account the efficiency of the time-amplitude analysis and the number of
accidental coincidences (3 pairs from 218 selected), the activity
of $^{228}$Th ($^{232}$Th family) inside $^{116}$CdWO$_4$ crystals
was determined to be 38(3) $\mu $Bq/kg \cite{Cd-00}.

The same technique was applied to the sequence of $\alpha $ decays from the
$^{235}$U family, and yields 5.5(14) $\mu $Bq/kg for $^{227}$Ac impurity in
the crystals.

The pulse shape (PS) of $^{116}$CdWO$_4$ events ($0.25$--$5$ MeV) is
digitized by a 12-bit ADC and stored in 2048 channels with 50 ns/channel
width. Due to different shapes of the scintillation signal for various kinds of
sources the PS technique based on the optimal digital filter was developed,
and clear discrimination between $\gamma $ rays and $\alpha $ particles was
achieved \cite{Faz98}. In the energy region 4.5-- 6 MeV for $\alpha $
particles (or 0.8--1.2 MeV for $\gamma $ quanta) numerical characteristics
of the shape (shape indicator, $SI$) are as following: $SI_\gamma $= 21.3 $%
\pm $ 2.0, and $SI_\alpha $= 32.5 $\pm $ 2.9. The PS selection technique
ensures the possibility to discriminate ''illegal'' events: double pulses, $%
\alpha $ events, etc., and thus suppress the background. For instance, PS
selection of the background events, whose $SI$ lie in the interval $%
SI_\gamma +2.4\sigma _\gamma <$ $SI$ $<$ $SI_\alpha +2.4\sigma _\alpha $ ($%
\approx $90\% of $\alpha $ events) yields the total $\alpha $ activity of $%
^{116}$CdWO$_4$ crystals as 1.4(3) mBq/kg. The last value can be adjusted
with activities determined by the time-amplitude analysis under the usual
assumption that secular radioactive equilibriums in some chains of the
$^{232}$%
Th and $^{238}$U families (e.g. $^{230}$Th $\rightarrow $ $^{226}$Ra
chain) are broken.

Since $SI$ characterizes the full signal, it is also useful to examine the
pulse front edge. It was found that at least 99\% of ''pure'' $\gamma $
events (measured with calibration $^{232}$Th source) satisfy the following
restriction on pulse rise time : $\Delta t$($\mu $s) $\leq $ 1.24 -- 0.5$%
\cdot E_\gamma $ + 0.078$\cdot E_\gamma ^2$, where $E_\gamma $ is in MeV.
Hence, the background pulses which do not pass this filter, were excluded
from the residual $\beta /\gamma $ spectrum.

The results of PS analysis of the data are presented in Fig. 4. The initial
(without PS selection) spectrum of the $^{116}$CdWO$_4$ scintillators in the
energy region $1.2$--$4$ MeV -- collected during 4629 h in anticoincidence
with the active shield -- is depicted in Fig. 4a, while the spectrum after PS
selection of the $\beta /\gamma $ events, whose $SI$ lie in the interval $%
SI_\gamma -3.0\sigma _\gamma \leq $ $SI\leq SI_\gamma +2.4\sigma _\gamma $
and $\Delta t$($\mu $s) $\leq $ 1.24 -- 0.5$\cdot E_\gamma $ + 0.078$\cdot
E_\gamma ^2$ (containing 98\% of $\beta /\gamma $ events), is shown in Fig.
4b. From these figures the background reduction due to pulse-shape analysis
is evident. Furthermore, Fig. 4c represents the difference between spectra in
Fig. 4a and 4b. These events, at least for energy above 2 MeV, can be
produced by $^{228}$Th activity from the intrinsic contamination of the $%
^{116}$CdWO$_4$ crystals (measured by the time-amplitude analysis as
described above). Indeed, two decays in the fast chain $^{212}$Bi ($Q_\beta $%
= $2.25$ MeV) $\rightarrow $ $^{212}$Po ($Q_\alpha $= $8.95$ MeV, $T_{1/2}$= 
$0.3$ $\mu $s) $\rightarrow $ $^{208}$Pb cannot be time resolved in the CdWO%
$_4$ scintillator (with an exponential decay time $\approx $15 $\mu $s \cite
{Geo96}, \cite{Faz98}) and will result in one event. To determine the
residual activity of $^{228}$Th in the crystals, the response function of the 
$^{116}$CdWO$_4$ detectors for the $^{212}$Bi $\rightarrow $ $^{212}$Po $%
\rightarrow $ $^{208}$Pb chain was simulated with GEANT3.21
and event generator DECAY4. The simulated function is shown in Fig.
4c, from which one can see that the high energy part of the experimental
spectrum is well reproduced ($\chi ^2$ = $1.3$) by the expected response for 
$^{212}$Bi $\rightarrow ^{212}$Po $\rightarrow ^{208}$Pb decays\footnote{%
The rest of the spectrum below 1.9 MeV (Fig. 4c) can be explained as the
high energy
tail of the PS selected $\alpha $ particles.}. Corresponding activity of $%
^{228}$Th inside the $^{116}$CdWO$_4$ crystals, deduced from the fit in the
1.9--3.7 MeV energy region, is 37(4) $\mu $Bq/kg, that is in a good
agreement with the value determined by the time-amplitude analysis of the
chain $^{220}$Rn $\rightarrow $ $^{216}$Po $\rightarrow $ $^{212}$Pb 38(3) $%
\mu $Bq/kg.

\nopagebreak
\begin{figure}[ht]
\begin{center}
\mbox{\epsfig{figure=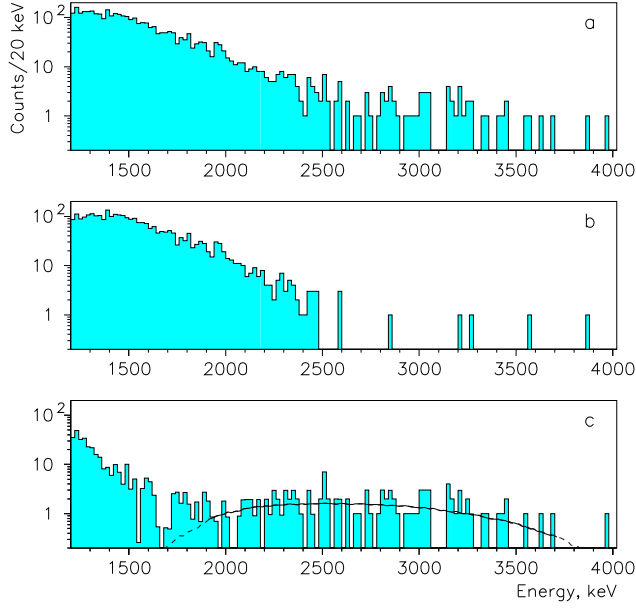,height=8.0cm}}
\caption {Initial spectrum of $^{116}$CdWO$_4$ crystals (339 g, 4629
h) without pulse-shape discrimination; (b) PS\ selected $\beta /\gamma $
events (see text); (c) the difference between spectra in Fig. 4a and 4b
together with the fit by the response function for $^{212}$Bi $\rightarrow $
$^{212}$Po $\rightarrow $ $^{208}$Pb decay chain. The fit value of $^{228}$Th
activity inside $^{116}$CdWO$_4$ crystals is 37(4) $\mu $Bq/kg.}
\end{center}
\end{figure}

To estimate the half-life limits for different neutrinoless $2\beta $ decay
mode, the simple background model was used. In fact, in the $0\nu 2\beta $
decay energy region only three background components (presented in fig. 3)
are important: (i) external $\gamma $ background from U/Th contamination of
the PMTs; (ii) the tail of the $2\nu 2\beta $ decay spectrum; and (iii) the
internal background distribution expected from the $^{212}$Bi $\rightarrow $ 
$^{212}$Po $\rightarrow $ $^{208}$Pb decay ($^{228}$Th chain). The limits
for the
neutrinoless mode of 2$\beta $ decay are set as $T_{1/2}$ $\geq 0.7(2.5$)%
$\cdot $1$0^{23}$ yr at 90\%(68\%) C.L. (for transition to the 
ground state of $%
^{116}$Sn), while for decays to the first 2$_1^{+}$ and second 0$_1^{+}$
excited levels of $^{116}$Sn as $T_{1/2}$ $\geq 1.3(4.8$)$\cdot $1$0^{22}$
yr and $\geq 0.7(2.4$)$\cdot $1$0^{22}$ yr at 90\%(68\%) C.L., accordingly.
For $0\nu $ decay with emission of one or two Majorons, the limits are: $%
T_{1/2}$(0$\nu $M1) $\geq 3.7(5.9$)$\cdot $1$0^{21}$ yr and $T_{1/2}$(0$\nu $%
M2) $\geq 5.9(9.4$)$\cdot $1$0^{20}$ yr at 90\%(68\%) C.L. Also the
half-life of $^{116}$Cd two neutrino $2\beta $ decay is determined as $%
T_{1/2}(2\nu 2\beta )=2.6\pm 0.1$(stat)$_{-0.4}^{+0.7}$(syst)$\cdot 10^{19}$
yr \cite{Cd-00}.

For illustration the high energy part of the experimental spectrum of the 
$^{116}$%
CdWO$_4$ detectors measured during 4629 h (histogram) is shown in Fig. 5
together with the fit by the 2$\nu $2$\beta $ contribution 
($T_{1/2}$=2.6$\cdot $%
1$0^{19}$ yr). The smooth curves $0\nu 2\beta $M1 and $0\nu 2\beta $M2 are
90\% C.L. exclusion distributions of M1 and M2 decays of $^{116}$Cd with 
$T_{1/2}$=$3.$7$\cdot $1$0^{21}$ yr and $T_{1/2}$= $5.$9$\cdot $1$0^{20}$
yr, respectively. In the insert, the peak from $0\nu 2\beta $ decay with $%
T_{1/2}$(0$\nu )$ = $1.$0$\cdot $10$^{22}$ yr and 90\% C.L. exclusion
(solid histogram) with $T_{1/2}$(0$\nu )$ = $7.$0$\cdot $1$%
0^{22}$ yr are depicted for comparison.

\nopagebreak
\begin{figure}[ht]
\begin{center}
\mbox{\epsfig{figure=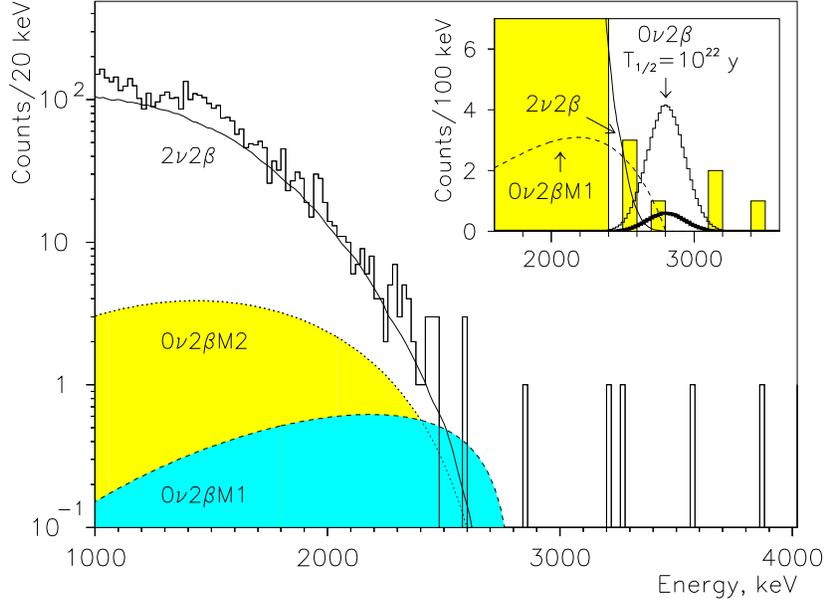,height=8.0cm}}
\caption {Part of the $^{116}$CdWO$_4$ spectrum measured for 4629 h
(histogram) and the fit by 2$\nu $2$\beta $ contribution ($T_{1/2}$ = 2.6$%
\cdot $1$0^{19}$ yr). The smooth curves are excluded (90\% C.L.)
distributions of 0$\nu $M1 and 0$\nu $M2 2$\beta $ decay of $^{116}$Cd with $%
T_{1/2}$ = $3.$7$\cdot $1$0^{21}$ yr and $T_{1/2}$ = $5.$9$\cdot $1$0^{20}$
yr, correspondingly. In the insert the peak from $0\nu 2\beta $ decay with $%
T_{1/2}$(0$\nu )$ = $1.$0$\cdot $10$^{22}$ yr is shown together with the
excluded (90\% C.L.) distribution (solid histogram) with $T_{1/2}$(0$\nu )$
= $7.$0$\cdot $1$0^{22}$ yr.}
\end{center}
\end{figure}

The following restrictions on the neutrino mass (using calculations \cite
{Klap90}) and neutrino-Majoron coupling constant (on the basis of
calculation \cite{Hir96b}) are derived from the experimental results
obtained: $%
m_\nu \leq 2.6(1.4)$ eV and $g_M\leq 12(9.5$)$\cdot $1$0^{-5}$ at 90\%(68\%)
C.L. \cite{Cd-00}. It is expected that after $\approx $5 years of
measurements a neutrino mass limit of $m_\nu \leq 1.2$ eV would be found.
However further advance of this limit to the sub-eV neutrino mass
domain is impossible without substantial sensitivity enhancement, which can
be reached with a greater number of $^{116}$CdWO$_4$ detectors ($\approx $65 kg)
placed in the liquid scintillator of the CTF.

\subsection{Design concept of the CAMEO-II project with $^{116}$Cd}

In the preliminary design concept of the CAMEO-II experiment, 24 enriched $%
^{116}$CdWO$_4$ crystals of large volume ($\approx $350 cm$^3$) are
located in the liquid scintillator of the CTF and fixed at 0.4 m
distance from the CTF center, thus homogeneously spread out on a sphere
with diameter 0.8 m. With a mass of 2.7 kg 
per crystal ($\oslash $7$\times $9 cm) the total $^{116}$Cd mass is 20
kg ($\approx $10$^{26}$ nuclei of $^{116}$Cd). It is
proposed that 200 PMTs with light concentrators are fixed at diameter 5 m
providing the total optical coverage of 80\% (see footnote 3).

The light output of CdWO$_4$ scintillator is (35--40)\% of NaI(Tl) which
yields $\approx $1.5$\cdot $10$^4$ emitted photons per MeV of energy
deposited. With a total light collection of $\approx $80\%
and PMT quantum efficiency of $\approx $25\% energy resolution of
several \% at 1 MeV can be obtained.

To justify this value a GEANT Monte Carlo simulation of the light
propagation in this geometry was performed, which gave $\approx $4000
photoelectrons for 2.8 MeV energy deposit. Therefore, with total optical
coverage 80\% the neutrinoless $2\beta $ decay peak of $^{116}$Cd ($Q_{\beta
\beta }$ $=$ 2805 keV) can be measured with energy resolution $FWHM$ =
4\%.
The principal feasibility to obtain such an energy resolution with CdWO$_4$
crystal situated in a liquid has been successfully demonstrated by measurements
with the help of a simple device. A cylindrical CdWO$_4$ crystal (40 mm in 
diameter and 30 mm in height) was fixed in the centre of a teflon container
with inner diameter 70 mm. This was coupled on opposite sides with 
two PMTs Philips XP2412, so that the distance from each flat surface of crystal
to the corresponding PMT's photocathode is 30 mm, while the gap between
the side surface of the crystal and the inner surface of the container is 15 mm. 
The container was filled up with the pure and transparent paraffin oil
(refractive index $\approx$1.5). Two PMTs work in coincidence and
results of measurements with $^{207}$Bi source are depicted in Fig. 6, where
the spectrum obtained with the standard detector arrangement (CdWO$_4$
crystal wrapped by teflon diffuser and directly coupled to the PMT's
photocathode with optical glue) is also shown for comparison.
As evident from Fig. 6, a substantial ($\approx$42\%) increase
in light collection was obtained with CdWO$_4$ in the liquid.
It resulted in improvement of the detector energy resolution in the whole
energy region 50--3000 keV (see Fig. 7 where the spectra measured with
$^{137}$Cs, $^{60}$Co and $^{232}$Th sources are presented). It should be
stressed that $FWHM$ values (7.4\% at 662 keV; 5.8\% at 1064 keV;
5.4\% at 1173 keV and 4.3\% at 2615 keV) are similar to those for NaI(Tl)
crystals and have never been reached before with CdWO$_4$ crystal scintillators
\cite{Geo96}.

\nopagebreak
\begin{figure}[ht]
\begin{center}
\mbox{\epsfig{figure=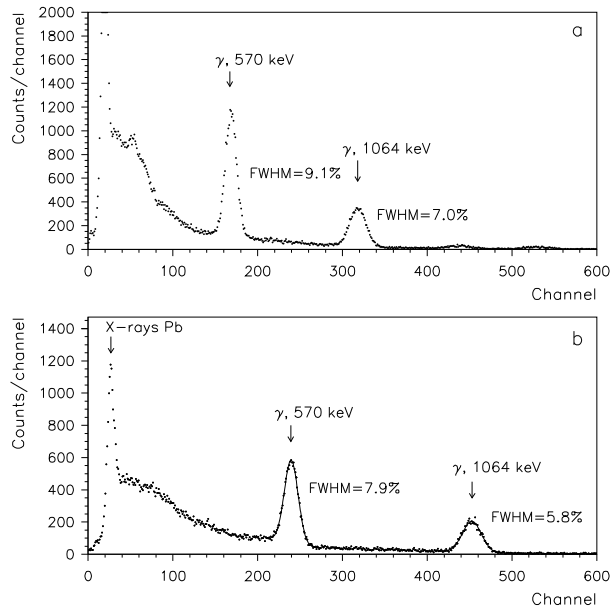,height=8.0cm}}
\caption {The energy spectra of a $^{207}$Bi source measured by a CdWO$_4$
crystal ($\oslash 40\times 30$ mm) for two arrangements: (a) standard,
where the CdWO$_4$ crystal wrapped by teflon diffuser is directly coupled to 
the PMT's photocathode with optical glue;
(b) the CdWO$_4$ crystal is located in liquid and viewed by two distant
PMTs (see text).}
\end{center}
\end{figure}

\nopagebreak
\begin{figure}[ht]
\begin{center}
\mbox{\epsfig{figure=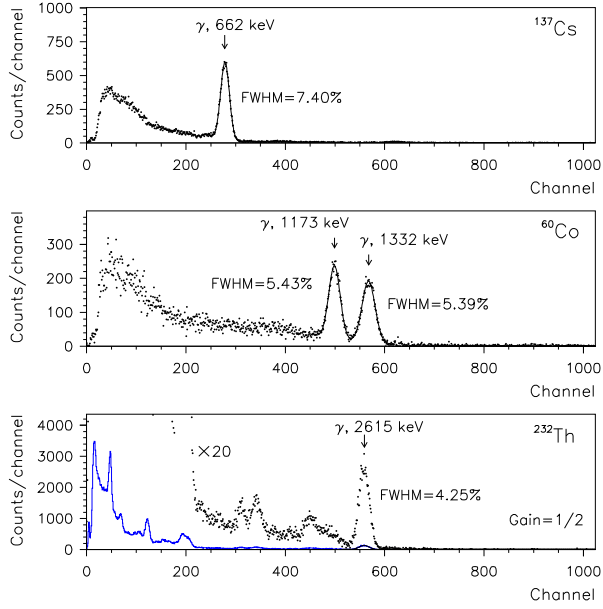,height=8.0cm}}
\caption {The energy spectra of $^{137}$Cs, $^{60}$Co and
$^{232}$Th sources measured by a CdWO$_4$ crystal ($\oslash 40\times 30$ mm)
placed in liquid and viewed by two distant PMTs.}
\end{center}
\end{figure}

Moreover, a strong dependence of the light collected by each individual PMT
versus position of the emitting source in the crystal was found. Such
a dependence can be explained by the large difference of the refraction
indexes of CdWO$_4$ crystal ($n=2.3$) and liquid scintillator ($n^{\prime
}=1.51$), which leads to light redistribution between reflection and
refraction processes due to changes of the source's position.

General formulae for the angular distribution
of the light emitted in the crystal and propagating in the liquid
scintillator are quite cumbersome, and below we give expressions for some
simple cases, when a CdWO$_4$ crystal with equal diameter and height 
($d=h=2a$) is placed in the center of the CTF detector
and a light source is positioned on the crystal axis. Assuming
a ratio of refraction indexes $n/n^{\prime }$ = $\sqrt{2}$ (which is close
to the real value) and neglecting light absorption, the equation
for the particular case with light source in the center of crystal is
of the form:

\begin{equation}
\frac{dW}{d\cos \theta }=\frac 1{2\sqrt{2}}\cdot [\frac{\left| \cos \theta
\right| }{\sqrt{1+\cos ^2\theta }}+1],
\end{equation}
where $\theta $ is the angle between the axis of the crystal (z-axis in our
coordinate system) and direction of the photon in the liquid
scintillator. This function is depicted in Fig. 8a together with the 
distribution for another case with the light source on the bottom of
crystal (on the axis again): 
\begin{equation}
\frac{dW}{d\cos \theta }=\frac 1{2\sqrt{2}}\cdot [\frac{\left| \cos \theta
\right| }{\sqrt{1+\cos ^2\theta }}+2\Theta (\theta )],
\end{equation}
where $\Theta (\theta )$ is the unit step function depending on angle $\theta $,
as following:

\begin{center}
$\Theta (\theta )=\left\{ 
\begin{array}{c}
1,\qquad 0\leq \theta \leq \pi /2 \\ 
0,\qquad \pi /2<\theta \leq \pi
\end{array}
.\right. $\\
\end{center}

\nopagebreak
\begin{figure}[ht]
\begin{center}
\mbox{\epsfig{figure=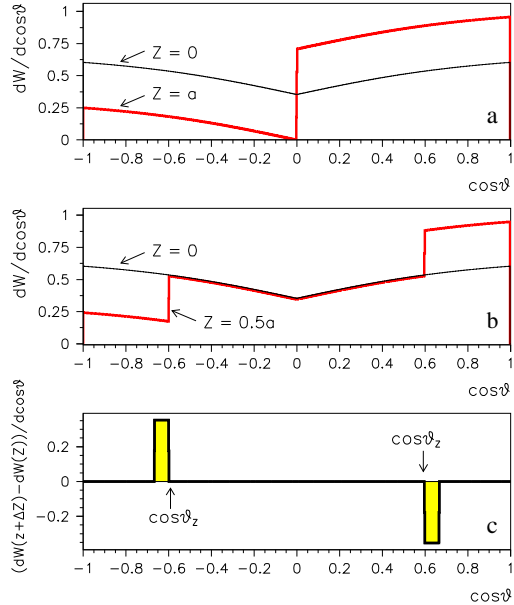,height=8.0cm}}
\caption {(a) Angular distribution of the light emitted in the CdWO$_4$
crystal and propagating in the liquid scintillator, when a light source is
on the crystal axis in the center and on the bottom of the crystal (solid
line). (b) The same as (a) but with a light source positioned on an
arbitrary point of the axis with coordinate $z$. (c) The difference between
angular distributions for two points on the axis with coordinates $z$ and $%
(z $ $+$ $\triangle z)$.arbitrary point of the axis with coordinate $z$. (c) The difference between
angular distributions for two points on the axis with coordinates $z$ and $%
(z $ $+$ $\triangle z)$.}
\end{center}
\end{figure}

The angular distribution for a more general case, when the light source is
positioned on an arbitrary point of the axis with coordinate $z$ ($-a\leq
z\leq a$), is shown in Fig. 8b. When the location of the source is shifted from $%
z$ to $z+\triangle z$, the values of angles $\theta _z$ and 
$\pi - \theta _z$,
for which $dW/d\cos \theta $ changes sharply, are also changed: $\cos
\theta _z=\sqrt{2(a-\left| z\right| )^2/(a^2+(a-\left| z\right| )^2)}$. The
difference $dW(z+\triangle z)/d\cos \theta -dW(z)/d\cos \theta $ represented
in Fig. 8c behaves as two peaks. The area of both peaks is equal to $\triangle
S=a^2\triangle z/\left[ a^2+(a-\left| z\right| )^2\right] ^{3/2}$. Assuming
that $N$ is the full number of photoelectrons detected by all PMTs, and $%
\triangle N$ is the difference of the number of photoelectrons due to the
source shift $\triangle z$, we find $\triangle N=N\cdot a^2\triangle
z/\left[ a^2+(a-\left| z\right| )^2\right] ^{3/2}$. From the last expression
one gets the formula for the spatial resolution in the CdWO$_4$ crystal
by supposing that difference $\triangle N$, which can be registered with
95\% C.L., is equal $\triangle N=2\sqrt{N}$:

\begin{center}
\begin{equation}
\triangle z=\frac{2a}{\sqrt{N}}\cdot \left\{ 1+\left[ 1-\frac{\left|
z\right| }a\right] ^2\right\} ^{3/2}.
\end{equation}
\end{center}

Substituting in Eq. (5) the crystal's dimensions $a=4$ cm, and the total number
of photoelectrons $N=4000$ yields spatial resolution of $\approx 4$ mm in
the center of the crystal ($z=0$) and $\approx 1.5$ mm near the top or
bottom of the crystal ($z=\pm a$). Our GEANT Monte Carlo simulation proves
these values and shows that, with a cylindrical CdWO$_4$ crystal ($\oslash
7\times 9$ cm) viewed by 200 PMTs, spatial resolution of 1--5 mm can be
reached depending on the event's location and the energy deposited in the
crystal (see, however, footnote 4).
The simulated distributions of the photoelectron number among PMTs due to
scintillations in CdWO$_4$ crystal are depicted in Fig. 9. The distance
between the source's positions on the crystal axis is equal to 
$\triangle z$ = 2.5
mm (Fig. 9a), and $\triangle z$ = 1.5 cm (Fig. 9b).

\nopagebreak
\begin{figure}[ht]
\begin{center}
\mbox{\epsfig{figure=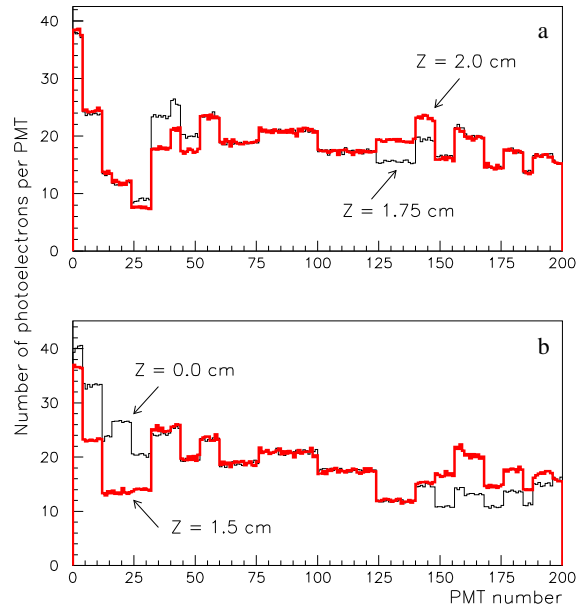,height=8.0cm}}
\caption {Simulated distributions of the photoelectron numbers among
PMTs due to scintillations in CdWO$_4$ crystal ($\oslash 7\times 9$ cm)
placed in the CTF. The distance between source positions on the crystal axis
equals $\triangle z$ = 2.5 mm (a), and $\triangle z$ = 1.5 cm (b).}
\end{center}
\end{figure}

These interesting features of light collection from $^{116}$CdWO$_4$ in the
CTF would allow a reduction in the contribution from high energy $%
\gamma $ quanta (e.g. $^{208}$Tl) to background in the energy region of
interest.
Besides, the non-uniformity of light collection can be
accurately corrected by using spatial information about each event 
in the CdWO$_4$ crystal, hence, helping to reach the required energy resolution
of the detectors.

\subsection{Background simulation}

The background simulations for the CAMEO-II experiment with $^{116}$Cd were
performed with the GEANT code and event generator DECAY4, and by
using the model described above.

\subsubsection{Two neutrino 2$\beta $ decays of $^{116}$Cd}

The half-life of two neutrino 2$\beta $ decay of $^{116}$Cd has been 
measured as $\approx $2.7$\cdot $10$^{19}$ yr \cite{Ejiri95,Dan95,Cd-00}.
The response functions of the CAMEO-II set up for $2\nu 2\beta $ decay of $%
^{116}$Cd with $T_{1/2}$ $=$ 2.7$\cdot $10$^{19}$ yr, as well as for $0\nu
2\beta $ decay with $T_{1/2}$ $=$ 10$^{25}$ yr were simulated, and results
are depicted in Fig. 2b. The calculated values of efficiency for the
neutrinoless channel are 86.1\% (for energy window 2.7 -- 2.9 MeV), and
75.3\% (2.75 -- 2.9 MeV). Background in the corresponding energy interval
from $2\nu 2\beta $ decay distribution is 2.3 counts/yr (2.7 -- 2.9 MeV) or
0.29 counts/yr (2.75 -- 2.9 MeV).

\subsubsection{Radioactive contamination of $^{116}$CdWO$_4$ crystals}

The very low levels of radioactive impurities by $^{40}$K and nuclides from
natural radioactive chains of $^{232}$Th and $^{238}$U in the enriched and
natural CdWO$_4$ crystals were demonstrated by the INR (Kiev) experiment 
\cite{Geo96}. On this basis the contamination criterion for $^{214}$Bi and $%
^{208}$Tl has been accepted in our calculation as $\approx $10 $\mu $Bq/kg,
which is equal to the actual activity value or limit determined for different
samples of CdWO$_4$ crystals \cite{Geo96}.

The calculated background contribution from the sum of $^{208}$Tl and $^{214}$%
Bi activities is $\approx $2000 counts/yr in the energy interval 2.7 -- 2.9
MeV. However, applying the time-amplitude analysis with spatial
resolution and pulse-shape discrimination technique developed 
\cite{Dan95,Cd-00} this
background rate can be reduced to $\approx $0.2 counts/yr or $\approx $1.0
counts for 5 years measuring period.

\subsubsection{Cosmogenic activities in $^{116}$CdWO$_4$}

For the CAMEO-II project cosmogenic activities in the 
$^{116}$CdWO$_4$ detectors
were calculated by the program COSMO \cite{Mar92}. A 1 month
exposure period on the Earth's surface was assumed 
and a deactivation time of about three
years in the underground laboratory. Only two nuclides produce background
in the energy window of the $^{116}$Cd neutrinoless $2\beta $ decay. These
are $^{110m}$Ag ($Q_\beta $=3.0 MeV; $T_{1/2}$=250 d) and $^{106}$Ru ($%
Q_\beta \approx $ 40 keV; $T_{1/2}$=374 d) $\rightarrow $ $^{106}$Rh ($%
Q_\beta $=3.5 MeV; $T_{1/2}$=30 s). Fortunately $^{106}$Ru activity is low
and the time-amplitude analysis can be applied ($T_{1/2}$=30 s), its estimated
background is practically negligible: $\approx $0.1 counts/yr in the energy
region 2.7 -- 2.9 MeV, and 0.05 counts/yr (2.75 -- 2.9 MeV). The 
background from 
$^{110m}$Ag is quite large: $\approx $23 (or $\approx $20) counts/yr for the
energy interval 2.7 -- 2.9 MeV (2.75 -- 2.9 MeV). However its contribution
can be reduced significantly by using spatial information because $^{110m}$%
Ag decays are accompanied by cascades of $\gamma $ quanta with energies $\geq 
$ 600 keV, which would be absorbed in spatially separated parts of the detector
giving an anticoincidence signature. Simulation under the assumption that the $%
^{116}$CdWO$_4$ crystal consists of small independent detectors with $%
h=d=1.2$ cm, yields the residual background rates $\approx $0.3 (or 0.2)
counts/yr in the corresponding energy region 2.7 -- 2.9 MeV (2.75 -- 2.9
MeV). The simulated spectrum from the cosmogenic activity of $^{110m}$Ag is
depicted in Fig. 2b.

\subsubsection{External background}

As in the previous case with $^{100}$Mo from the various sources of external
background only $\gamma $ quanta due to PMT contamination and from Rn
impurities in the shielding water were simulated, while others were simply
estimated as negligible on the basis of the results of ref. \cite{Dark98}.
The radioactivity values of the EMI 9351 PMT accepted for the simulation
are: 0.194 Bq/PMT for $^{208}$Tl; 1.383 Bq/PMT for $^{214}$Bi; and 191
Bq/PMT for $^{40}$K \cite{CTF-98A,Xe-CTF}. Also possible $^{222}$Rn activity
in the shielding water (in the region surrounding the inner vessel) at the
level of $\approx $30 mBq/m$^3$ was taken into account. The simulation
performed under these assumptions finds that the only important
contribution to the background in the vicinity of the $0\nu 2\beta $ decay
energy is $^{208}$Tl activity from the PMTs. The calculated values are $\approx 
$0.8 and 0.05 counts/yr in the energy interval 2.7 -- 2.9 MeV (2.75 -- 2.9
MeV). However, with the help of spatial information available for each event
occurring inside the 
$^{116}$CdWO$_4$ crystal, these contributions can be reduced
further to the level of $\approx $0.08 (or 0.005) counts/yr in the energy
region 2.7 -- 2.9 MeV (2.75 -- 2.9 MeV). The simulated background
contribution from $^{208}$Tl contamination of the PMTs is shown in Fig. 2b.
Summarizing all background sources gives $\approx $3 counts/yr (0.6
counts/yr) in the energy interval 2.7 -- 2.9 MeV (2.75 -- 2.9 MeV).

\subsection{Sensitivity of the $^{116}$Cd experiment}

As earlier we will estimate the sensitivity of the CAMEO-II experiment with the
help of Eq. (2). Taking into account the number of $^{116}$Cd nuclei ($%
\approx $10$^{26}$), measuring time of 5--8 years, detection efficiency of
75\%, and with expected background of 3--4 counts, one can obtain a half-life
limit $T_{1/2}$(0$\nu 2\beta $) $\geq $ 10$^{26}$ yr. On the other hand,
it is evident from Fig. 2b that neutrinoless $2\beta $ decay of $^{116}$Cd
with half-life of $\approx $10$^{25}$ yr would be clearly registered.

It should be stressed that such a level of sensitivity for 0$\nu 2\beta $
decay cannot be reached in the presently running $2\beta $ decay experiments
(perhaps only with $^{76}$Ge), as well as for approved projects, like NEMO-3 
\cite{NEMO98} and CUORICINO \cite{CUORE98}, which are under construction
now. It was shown above that the sensitivity of the
NEMO-3 set up is limited at the
level of $\approx $4$\cdot $10$^{24}$ yr by the detection efficiency and
energy resolution (see Fig. 1f). 
The CUORICINO project is designed
to study $2\beta $ decay of $^{130}$Te with the help of 60 low temperature
bolometers made of TeO$_2$ crystals (750 g mass each). Another aim of 
CUORICINO is to be a pilot step for a future CUORE project (not approved
yet), which would consists of one thousand TeO$_2$ bolometers with total
mass of 750 kg \cite{CUORE98}. Despite the excellent energy resolution of
these detectors ($\approx $5 keV at 2.5 MeV) the main disadvantage of the
cryogenic technique is its complexity, which requires the use of a lot of
different construction materials in the apparatus. This fact, together with
the lower $2\beta $ decay energy of $^{130}$Te ($Q_{\beta \beta }$=2528 keV),
makes it quite difficult to reach the same super-low level of background
as obtained in experiments with semiconductor and scintillation
detectors. For example, one can compare the background rate of 0.6 counts/yr$%
\cdot $kg$\cdot $keV at 2.5 MeV from the current Milano experiment with
twenty TeO$_2$ bolometers \cite{Te130,CUORE98} with the 
value of 0.03 counts/yr$%
\cdot $kg$\cdot $keV in the energy region $2.5$--$3.2$ MeV, which was reached
in the Kiev-Firenze experiment with $^{116}$CdWO$_4$ crystal scintillators 
\cite{Cd-00}. In that sense the CAMEO-II experiment 
has a great fundamental advantage
because signaling from the $^{116}$CdWO$_4$ crystals to the
PMTs (placed far away
from crystals) is provided by light propagating in the super-low background
medium of liquid scintillator, whereas cryogenic or semiconductor detectors
must be connected with receiving modules by cables. These
additional materials (wires, insulators, etc.), whose radioactive
contamination are much larger in comparison with TeO$_2$ crystals, Ge
detectors or liquid scintillators, must be introduced in the neighborhood
of the main detectors, giving rise to additional background\footnote{%
There are two origins of such a background: i) radioactive contamination of
the materials introduced; ii) external background penetrating through the slots
in the detector shielding required for the connecting cables.}.

Another drawback of cryogenic detectors is their low reliability.
At the same time, the CAMEO-II
experiment with $^{116}$CdWO$_4$ crystals is simple and reliable,
and therefore can run for decades without problems and with very low maintenance
cost\footnote{%
It should be noted that the $^{116}$CdWO$_4$ crystals produced for the CAMEO-II
experiment can also be used as cryogenic detectors with high energy
resolution \cite{CdWO-94}. In the event of a positive effect seen by CAMEO-II 
these crystals could be measured by the CUORE apparatus; in some
sense both projects are complementary.}.

Moreover, the CAMEO-II project can be advanced farther by exploiting one ton of $%
^{116}$CdWO$_4$ detectors ($\approx $1.5$\cdot 1$0$^{27}$ nuclei of $^{116}$%
Cd) and the BOREXINO apparatus (CAMEO-III). With this aim 370 enriched $%
^{116}$CdWO$_4$ crystals (2.7 kg mass of each) would be placed at a diameter
3.2 m in the BOREXINO liquid scintillator. The simulated
response functions of such a detector system for $2\nu 2\beta $ decay of $%
^{116}$Cd with $T_{1/2}$ $=$ 2.7$\cdot $10$^{19}$ yr, as well as for $0\nu
2\beta $ decay with $T_{1/2}$ $=$ 10$^{26}$ yr considering a 10-year measuring
period are depicted in Fig. 2c. Because background in BOREXINO 
should be even lower than in the CTF, the sensitivity of CAMEO-III
for neutrinoless $2\beta $ decay of $^{116}$Cd is estimated as $%
T_{1/2}$ $\geq $ 10$^{27}$ yr, while $0\nu 2\beta $ decay with
half-life of $\approx $10$^{26}$ yr can be detected. This level of
sensitivity can be compared only with that of the GENIUS project \cite
{GENIUS-98}, which is under discussion now and intends to operate one ton of
Ge (enriched in $^{76}$Ge) semiconductor detectors placed in a tank ($%
\oslash 12\times 12$ m) with extremely high-purity liquid nitrogen
(required demands on its radioactive contamination are $\approx 10^{-15}$
g/g for $^{40}$K, $^{238}$U, $^{232}$Th, and 0.05 mBq/m$^3$ for $^{222}$Rn 
\cite{GENIUS-98}) serving as cooling medium and shielding for the detectors
simultaneously. Let us estimate the sensitivity of GENIUS in the same way as
that for CAMEO. One ton of Ge detectors with enrichment $\approx $86\% (as
in the current $^{76}$Ge experiments \cite{IGEX,Ge76}) would provide $\approx $7$%
\cdot $10$^{27}$ nuclei of $^{76}$Ge, thus in the optimistic case of zero
background a sensitivity of $T_{1/2}^{0\nu }$ $\geq $ 5$\cdot $10$^{27}$
yr can be reached there.

By the aid of Eq. (1) one can obtain an expression for the neutrino mass bound
derived from the experimental half-life limit for 0$\nu 2\beta $ decay as
$\lim \left\langle m_\nu \right\rangle = \left\{ \lim
T_{1/2}^{0\nu }\cdot G_{mm}^{0\nu }\cdot \left| ME\right| ^2\right\}
^{-1/2}. $ Because of the lower $2\beta $ decay energy the phase-space
integral $G_{mm}^{0\nu }$ for 0$\nu 2\beta $ decay of $^{76}$Ge is about ten
times lower than for $^{116}$Cd. Hence, it is evident from the last equation
that neglecting the complicated problem of nuclear matrix elements
calculation, the CAMEO-III experiment will bring at least the same restriction
on the neutrino mass as the GENIUS project. Indeed, on the basis of the CAMEO
half-life limit $T_{1/2}^{0\nu }$ $\geq $ 10$^{27}$ yr and using
calculations \cite{Klap90,Arn96} one can derive a limit on the neutrino
mass of $\approx 0.02$ eV at 90\% C.L., which is practically equal to the
value $\approx 0.01$ eV claimed as the main goal of GENIUS \cite
{GENIUS-98}. At the same time it is obvious that the technical tasks, whose
solutions are required for the realization of these super--high sensitive
projects (GENIUS, CUORE, and CAMEO) are simpler for CAMEO. In fact, the
super-low background apparatus needed for the last experiment is already
running (this is the CTF) or under construction (BOREXINO),
while for the CUORE and GENIUS proposals such unique apparati should be designed
and constructed.

\section{Conclusions}

1. The unique features of the CTF and BOREXINO (super-low background
and large sensitive volume) are used to develop a realistic, competitive,
and efficient program for high sensitivity 2$\beta $ decay research (CAMEO\
project). This program includes three natural steps, and each of them would
bring substantial physical results:

\hspace{1.0in}

{\it CAMEO-I.}{\bf \ }With a passive 1 kg source made of $^{100}$Mo ($^{116}$%
Cd, $^{82}$Se, $^{150}$Nd) and located in the liquid scintillator of the
CTF, the sensitivity (in terms of the half-life limit for $0\nu 2\beta $
decay) is (3--5)$\cdot $10$^{24}$ yr. It corresponds to a bound on the
neutrino mass $m_\nu \leq $ 0.1--0.3 eV, which is similar to or better than
those of running ($^{76}$Ge), and future NEMO-3 ($^{100}$Mo) and CUORICINO (%
$^{130}$Te) experiments.

\hspace{1.0in}

{\it CAMEO-II.} With 24 enriched $^{116}$CdWO$_4$ crystal scintillators
(total mass of 65 kg) placed as ''active'' detectors in the liquid
scintillator of the CTF the sensitivity would be $\approx $10$^{26}$ yr.
Such a half-life limit could be obtained only 
by future CUORE ($^{130}$Te) and GENIUS ($^{76}$Ge) projects. 
Pilot $^{116}$Cd research, performed by the INR (Kiev)
during the last decade, as well as Monte Carlo simulation show the
feasibility of CAMEO-II, which will yield a limit on the
neutrino mass $m_\nu \leq $ 0.05--0.07 eV.

\hspace{1.0in}

{\it CAMEO-III.}{\bf \ }By exploiting one ton of $^{116}$CdWO$_4$ detectors
(370 enriched $^{116}$CdWO$_4$ crystals) introduced in the BOREXINO liquid
scintillator, the half-life limit can be advanced to the level of $\approx $%
10$^{27}$ yr, corresponding to a neutrino mass bound of $\approx 0.02$ eV.

\hspace{1.0in}

2. In contrast to other projects CAMEO has three principal advantages:

i) Practical realization of the CAMEO project is simpler due to the use of
already existing super-low background CTF or (presently under construction)
BOREXINO apparatus;

ii) Signaling from the $^{116}$CdWO$_4$ crystals to PMT (placed far away) is
provided by light propagating in the high-purity medium of liquid
scintillator -- this allows practically zero background to be reached
in the energy region of the 0$\nu 2\beta $ decay peak;

iii) Extreme simplicity of the technique used for 2$\beta $ decay study 
leads to high reliability and low maintenance costs for the CAMEO
experiment, which therefore can run permanently and stably for decades.

\hspace{1.0in}

3. Fulfillment of the CAMEO program would be a real breakthrough in the
field of $2\beta $ decay investigation, and will bring outstanding results
for particle physics, cosmology and astrophysics. Discovery of neutrinoless $%
2\beta $ decay will clearly and unambiguously manifest new physical
effects beyond the Standard Model. In the event of a null result, the limits
obtained by the CAMEO experiments would yield strong restrictions on parameters
of manifold extensions of the SM (neutrino mass and models; dark matter and
solar neutrinos; right-handed contributions to weak interactions; leptoquark
masses; bounds for parameter space of SUSY models; neutrino-Majoron coupling
constant; composite heavy neutrinos; Lorentz invariance, etc.), which
will help to advance basic theory and our understanding of the origin and
evolution of the Universe.

\end{document}